\documentclass[onecolumn,10.5pt,compsoc]{astro}

\usepackage[bf]{caption}
\usepackage{cuted}  
\usepackage{epstopdf}
\usepackage{enumitem}
\usepackage{mathtools}
\DeclarePairedDelimiter\ceil{\lceil}{\rceil}
\DeclarePairedDelimiter\floor{\lfloor}{\rfloor}
\usepackage{bm}
\DeclareCaptionLabelSeparator{threespace}{~\,~}
\usepackage{longtable}

\begin{document}
\pretitle{}
\title{Target selection for a small low-thrust mission to near-Earth asteroids}
\author{Alessio Mereta and Dario Izzo \cor}
\addr{Advanced Concepts Team, European Space Research and Technology Center,
Keplerlaan 1, 2201 AZ Noordwijk, The Netherlands}

\coremail{Dario Izzo: dario.izzo@esa.int}
\abstracts{The preliminary mission design of spacecraft missions to asteroids often involves, in the early phases, the selection of candidate target asteroids. The final result of such an analysis is a list of asteroids, ranked with respect to the necessary propellant to be used, that the spacecraft could potentially reach. In this paper we investigate the sensitivity of the produced asteroids rank to the employed trajectory model in the specific case of a small low-thrust propelled spacecraft beginning its journey from the Sun-Earth $L_2$ Lagrangian point and heading to a rendezvous with some near-Earth asteroid. We consider five increasingly complex trajectory models: impulsive, Lambert, nuclear electric propulsion, nuclear electric propulsion including the Earth's gravity, solar electric propulsion including the Earth's gravity and we study the final correlation between the obtained target rankings. We find that the use of a low-thrust trajectory model is of great importance for target selection, since the use of chemical propulsion surrogates leads to favouring less attractive options 19\% of times, a percentage that drops to 8\% already using a simple nuclear electric propulsion model that neglects the Earth's gravity effects and thrust dependence on the solar distance. We also find that for the study case considered, a small interplanetary CubeSat named M-ARGO, the inclusion of the Earth's gravity in the considered dynamics does not affect the target selection significantly.}
\keywords{ low-thrust, asteroid selection, near-Earth asteroids, mission analysis} 

\section*{Nomenclature}
\begingroup
\vspace {-3mm}
\begin{center}
\renewcommand{\arraystretch}{1.}
\begin{longtable}{|p{40mm}|p{120mm}|}
\hline
OCP&
Optimal Control Problem \\
\hline
NLP&
Nonlinear Programming Problem \\
\hline
SEP&
Solar Electric Propulsion \\
\hline
NEP&
Nuclear Electric Propulsion \\
\hline
CDF&
Concurrent Design Facility\\
\hline
$t_1$&
launch epoch \\
\hline
$t_2$&
arrival epoch \\
\hline
$tof$&
time of flight \\
\hline
$T_{max}$&
maximum thrust (mN) \\
\hline
$I_{sp}$&
specific impulse (s) \\
\hline
$m, \bm{r}, \bm{v}$&
spacecraft mass (kg), position (m) and velocity vector (m/s) \\
\hline
$m_{i}$&
initial spacecraft mass (kg) \\
\hline
$m_{p}$&
propellant mass (kg) \\
\hline
$m_{f}$&
final spacecraft mass ($m_i-m_p$) (kg) \\
\hline
$r_S$&
Sun-spacecraft distance (m)\\
\hline
$r_E$&
Earth-spacecraft distance (m)\\
\hline
$\bm{r_E}$&
Earth-spacecraft vector (m)\\
\hline
$\bm{r_1}, \bm{v}_1$&
Earth position (m) and velocity vector (m/s) at $t_1$\\
\hline
$\bm{r_2}, \bm{v}_2$&
target asteroid position (m) and velocity vector (m/s) at $t_2$\\
\hline
$P_{in}, P_{sw}, P_{bmp}$&
power to the electric propulsion subsystem, from the solar panels and from the body mounted panel (watt)\\
\hline
$\Delta i_{rel}$&
inclination of the target body with respect to the departure body\\
\hline
$\mu, \mu_E$&
gravitational parameters for the Sun and the Earth (m$^3$/s$^2$)\\
\hline
$g_0$&
standard gravity on the Earth (m/s$^2$)\\
\hline
\end{longtable}
\label{tab1}
\end{center}
\vspace {-6mm}
\endgroup

\section{Introduction}
\noindent Low-thrust electric propulsion offers a higher efficiency compared to chemical propulsion, and is an option often considered for interplanetary missions. The Bepi Colombo mission \cite{bepi}, the Hyabusa mission \cite{hyabausa}, the Dawn mission \cite{dawn} as well as the SMART-1 mission \cite{smart1} are only a few important examples of actual mission designs that make use of a low-thrust propulsion system. Such propulsion systems provide very small amounts of thrust that must be applied over a significant fraction of the trajectory. This causes the optimal control problem (OCP) of transferring the spacecraft between bodies to be continuous rather than discrete, typically making the design of trajectories computationally more demanding. This is particularly problematic in preliminary mission design phases, when a large number of options need to be considered and studied, and in particular for the problem of identifying attractive targets (attractive from the dynamics point of view) for missions targeting one or more asteroids (e.g., Dawn~\cite{dawn}, Marco Polo~\cite{marcopolo}, Don Quijote~\cite{dq}). As a consequence, in preliminary phases the spacecraft propulsion system and the dynamics are often simplified as to be able to screen efficiently the many possible targets these missions may have, and help the final decision that will ultimately be made in later stages according to multiple criteria.  Purely impulsive models such as bi-impulsive Lambert transfers or Hohmann transfers as well as low thrusts simplified models such as the Edelbaum's approximation \cite{edelbaum} are often considered to produce a rapid target selection but, to the best of our knowledge, the error introduced by such simplifications on the final produced ranks has never been quantified. Such a quantification is the focus of this paper. Our work is motivated by an internal feasibility study on the small interplanetary CubeSat named M-ARGO. The study was performed in early 2017 at the European Space Agency's Concurrent Design Facility. M-ARGO is a small low-thrust mission propelled by solar electric propulsion, possibly starting from the Sun-Earth $L_2$ Lagrangian point and heading to a rendezvous with a near-Earth asteroid. For such a mission, during the preliminary mission design phase, the enumeration and ranking of all asteroids the spacecraft can reach was required and had to be produced. This, according to the model used for the spacecraft transfer, results in a list of asteroids that is to be further pruned accounting for other mission requirements before performing more mature trajectory designs - much more computationally intensive - on the few candidates that are left at the end of the process. Using M-ARGO as a study-case, we perform a detailed assessment on the accuracy that trajectory models have when used to assemble such preliminary target selection ranks.

The paper is organized as follows. In Section \S \ref{sec:methodology} we introduce the overall methodology we used in the study to asses the impact of different employed model on the resulting target selection. In Section \S \ref{sec:mission} we introduce the details on the selected case of the interplanetary CubeSat mission M-ARGO, including the capabilities of the solar electric propulsion subsystem. In Section \S \ref{sec:traj_models} we describe the five trajectory models considered in this study: a three impulses approximation, a Lambert optimal transfer, a nuclear electric propulsion two-body transfer, a nuclear electric propulsion transfer accounting for both the Sun and the Earth gravitational pulls and our ground truth model, i.e., a solar electric propulsion transfer accounting for both the Sun and the Earth gravitational pulls. In Section \S \ref{sec:results} we describe the experimental campaign and discuss its results in terms of the Kendall-tau rank correlation metric.

\section{Methodology}
\label{sec:methodology}
To assess the effect produced by the employed trajectory model on the final target ranking, we
consider five increasingly complex models. The first two are based on impulsive transfers, namely
a three impulses approximation and a multi-revolution Lambert approximation. To these we add three
low-thrust trajectory models we will refer to as NEP (nuclear electric propulsion), NEP+G
(nuclear electric propulsion plus gravity) and SEP+G (solar electric propulsion plus gravity).
We do not consider Edelbaum's approximation as in this multiple-revolution mass-optimal case it is not returning any accurate result.

In the NEP model the spacecraft thrust, as well as its specific impulse, is considered as fixed.
In the NEP+G model the Earth's gravity is added in the model since the spacecraft departure from L2
suggests it may play some role in the resulting transfer efficiency. Finally the SEP+G model is used as ground truth as it considers the complex thrust and specific impulse relation to the
spacecraft-Sun distance as well as the Earth and Sun gravity.
In the case of the low-thrust models the trajectory is optimized with respect to the final mass using the direct transcription method described in \cite{yam, sims}, extended
to also include the Earth's gravity and the SEP capability (in a similar way as the Sims-Flanagan model was extended in \cite{sepmalto}). The
code used is made available as part of the European Space Agency's \emph{pykep} project at
\url{https://goo.gl/u4dVBc}. Using each of the five models we optimize trajectories to each of 143 asteroids candidates
pre-filtered out of the whole catalogue of known minor planets \cite{mpcorb}.

Since each optimal transfer problem is, essentially, a multi-objective optimization problem where
not only the final mass $m_f$ but also the transfer time $tof$ is to be accounted for, we
preliminarily look at the non-dominated fronts and compare, for each of the targets, the result
obtained using the different low-thrust models. In the case of the impulsive models, the multi-objective
optimization perspective is not beneficial as the transfer times are not representative of an
actual low-thrust mission, hence they are not considered. The results indicate how, overall, the
transfer quality can be well represented by one single point on the non-dominated front, and we thus
fix the time of flight to belong to a small bin around $tof = 3$ years as that is also the requirement considered during the M-ARGO mission study.

Finally, we compute the propellant mass $m_p$ necessary to reach each asteroid according to each trajectory model and thus produce the target selection ranks in all cases. Kendall-tau statistics is then used to compare the rankings and, in particular, to find the probability to erroneously rank a randomly selected asteroid pair when using a given simplified model.

\section{Mission profile}
\label{sec:mission}
\noindent In this section we briefly describe the relevant parts of the mission profile considered for the M-ARGO mission as emerged from the internal study performed at the European Space Agency's Concurrent Design Facility. A small spacecraft, a 12U CubeSat, is starting at the Sun-Earth L2 Lagrangian point (reached, for instance, after piggybacking on another mission) and has to rendezvous with a near-Earth asteroid.  The departure epoch $t_1$ is bounded between the years 2020 and 2023 (included), and the maximum allowed time of flight $tof$ is 3 years. The spacecraft has an initial mass $m_i$ of 20 kg, including a propellant mass $m_p$ that, at most, can be 2.5 kg. Its propulsion system consists of a low-thrust engine powered by solar arrays. The final resulting model for the thrust can be expressed by the equations that follow.

\noindent The maximum thrust $T_{max} $ is given, in mN, by the expression:
$$ 
T_{max} = (26.27127 P_{in}-708.973) / 1000
$$
while the spacecraft specific impulse $ I_{sp}$, in s, is given by:
$$ 
I_{sp} = -0.0011 P_{in}^3 + 0.175971 P_{in}^2 + 4.193797 P_{in} + 2073.213
$$
with the input power $P_{in}$ to the electric propulsion subsystem given in W by:
$$ P_{in} = min\{\eta(P_{sw}-max\{13.75-P_{bmp},0\}), 120\} $$
where $\eta = 0.92$ is an efficiency ratio, and power comes from solar panels arranged in two separate buses:
\begin{itemize}
    \item a body-mounted solar panel that supplies only the platform bus with a power in W:
        $$ 
        P_{bmp} = -40.558r_S^3+173.49r_S^2-259.19r_S+141.86 
        $$
    \item solar wings that produce a power in W:
        $$ 
        P_{sw} = -146.26r_S^3+658.52r_S^2-1059.2r_S+648.24
        $$
\end{itemize}
where the Sun-spacecraft distance, in AU, is indicated by $r_S$. The platform requires a constant supply of 13.75 W, hence if $P_{bmp}$ is smaller we must subtract the remainder from $P_{sw}$. We also consider thermal constraints limiting the maximum power input to 120 W.

\begin{figure}[t]
    \center
    \includegraphics[width=.6\textwidth]{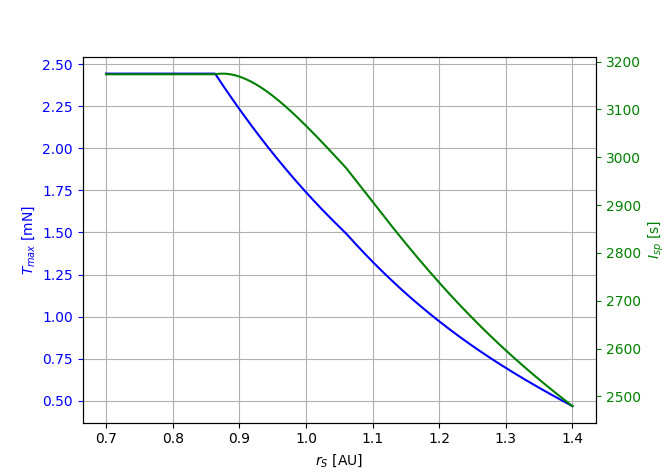}
    \caption{$T_{max}$ and $I_{sp}$ as a function of the distance from the Sun $r_S$.}
    \label{fig:M-ARGO_model}
\end{figure}

Fig.(\ref{fig:M-ARGO_model}) shows the resulting plots for $T_{max}$ and $I_{sp}$ as a function of the solar distance. Note that given the nature of the mission the part of the plot beyond 1.4 AU will not be used.

\section{Trajectory models}
\label{sec:traj_models}
\subsection{Three Impulses}
\label{ssec:imp}
\noindent In the trajectory model we call \lq\lq three impulses\rq\rq\ we approximate the $\Delta V$ necessary for an orbit transfer assuming only to match semi-major axis, eccentricity, inclination and right ascension of the ascending node (RAAN). The optimal transfer is, in this case, accomplished by three velocity changes: one to match the target aphelion, one to match the target perihelion and one to match the target inclination and RAAN. Two of the three impulses can be merged together at either departure (from the Earth) or arrival (at the target asteroid). Since the argument of perigee is not matched, nor is the mean anomaly, such a transfer is a good approximation of an impulsive transfer when one of the two eccentricities is small and the actual position along the orbit is not relevant. The resulting formula used to compute the required $\Delta V$ is as follows:
$$
\Delta V_{3I} = \Delta V_{dep} + \Delta V_{arr}
$$
where $\Delta V_{dep}$ is an initial velocity change delivered at the departure orbit and $\Delta V_{arr}$ is a final velocity change delivered at the arrival orbit. They are computed differently according to the values of the departure and arrival aphelion $r_{a1}, r_{a2}$:
\begin{enumerate}
\item $r_{a1} \le r_{a2}$: in this case the aphelion of the target orbit is larger than the
starting orbit aphelion. We thus perform an inclination change at arrival to match RAAN and
inclination, and we match the semi-major axis and eccentricity using a Hohmann-like transfer
from the perihelion of the starting orbit to the aphelion of the arrival orbit (we assume the
two arguments of perihelion to be identical):
\begin{eqnarray*}
\Delta V_{dep} / \sqrt\mu &=& \sqrt{2/r_{p_1}-2/(r_{p_1}+r_{a_2})} - \sqrt{2/r_{p_1}-2/(r_{p_1}+r_{a_1})} \\
V_i / \sqrt\mu &=& \sqrt{2/r_{a_2}-2/(r_{p_1}+r_{a_2})} \\
V_f / \sqrt\mu &=& \sqrt{2/r_{a_2}-2/(r_{p2}+r_{a2})} \\
\Delta V_{arr} &=& \sqrt{V_i^2+V_f^2-2V_iV_f \cos\Delta i_{rel}}
\end{eqnarray*}
with:
$$
\cos \Delta i_{rel} = \cos i_1\cos i_2 + \sin i_1\sin i_2\cos\Omega_1\cos\Omega_2+ \sin i_1\sin i_2\sin\Omega_1\sin\Omega_2
$$
where $r_{p1}$ and $r_{p2}$ are the departure and arrival perihelions, $V_i$ is the velocity of the transfer orbit, $V_f$ is the velocity of the arrival orbit, $i_1, i_2$ and $\Omega_1, \Omega_2$ are, respectively, the inclinations and RAANs of the departure and arrival orbits.
\item $r_{a1} > r_{a2}$: in this case the aphelion of the target orbit is smaller than the starting orbit aphelion. We thus perform an inclination change at departure to match RAAN and inclination and we match the semi-major axis and eccentricity using a Hohmann-like transfer from the aphelion of the starting orbit to the perihelion of the arrival orbit. To compute the $\Delta V_{3I}$ in this case, we can make use of the same equations reported, after having switched the departure and arrival orbits.
\end{enumerate}

\subsection{Lambert}
\label{ssec:lamb}
\noindent In the trajectory model we call \lq\lq Lambert\rq\rq\ the total $\Delta V$ is computed as the optimal two-impulse transfer between the Earth and the asteroid. Such a transfer is found by solving the optimization problem:
\begin{equation}
\label{eq:lambertP}
\mathcal P_1: 
\left\{
\begin{array}{ll}
\mbox{find:} & t_1 \in [\underline t, \overline t], tof \in [\underline{tof}, \overline{tof}]\\
\mbox{to minimize} & \Delta V_{L} \\
\end{array}
\right.
\end{equation}

where $t_1$ is the launch epoch and $\Delta V_{L}$ is the required velocity increment computed along the best multi-revolution (up to 5 revolutions) solution to the Lambert's problem of transferring in $tof$ from the Earth position $\bm{r_1}(t_1)$ to the target asteroid position $\bm{r_2}(t_1+tof)$. The algorithm we use to solve the Lambert's problem is described in \cite{revlamb}.
Such a relatively simple optimization is performed using an open source implementation of a Self-adaptive Differential Evolution algorithm \cite{pagmo2} (20 individuals, 500 generations). The bounds $[\underline t, \overline t]$ considered for the departure epoch are in the 2020-2023 window, while the bounds $[\underline{tof}, \overline{tof}]$ for the time of flight are between 0 and 3 years.

\subsection{NEP}
The trajectory model here called \lq\lq NEP\rq\rq\ stands for Nuclear Electric Propulsion model. It considers a spacecraft equipped with a low-thrust propulsion system able to deliver a constant, predefined thrust level $T_{max}$ at a constant, predefined specific
impulse $I_{sp}$. We fix $T_{max}$ = 1.7 mN and $I_{sp}$ = 3000 s.

In an inertial frame centered on the Sun, the equations of motion are given by:
\begin{eqnarray}
\label{eqn:mot}
\ddot{\bm{r}} & = & -\mu\frac{\bm{r}}{r_S^3}+\frac{\bm{u}(t)}{m}, \\
\dot{m} & = & -\frac{\left|\bm{u}(t)\right|}{I_{sp}g_0}, \nonumber
\end{eqnarray}
where $r_S = |\bm{r}|$, $\mu \approx 1.327\cdot10^{20}$ m$^3$/s$^2$ is the gravitational parameter of the Sun, $I_{sp}$ is the specific impulse of the low-thrust propulsion, and $g_0 \approx 9.8066$ m/s$^2$ is the standard gravity on Earth. The control $\bm{u}(t)$ is the thrust vector. We thus introduce the following optimal control problem (OCP):

\begin{equation}
\label{eq:OCP}
\mathcal P_2: 
\left\{
\begin{array}{ll}
\mbox{find:} & \bm u(t) \in \mathcal F, t_1 \in [\underline t, \overline t], tof \in [\underline{tof}, \overline{tof}] \\
\mbox{to maximize:} & J = m_f = m_i - \int_{t_1}^{t_2} \left|\bm{u}(t)\right|~dt \\
\mbox{subject to:} & \\
& \bm{r}(t_1)  =  \bm{r_1}, \\
&\dot{\bm{r}}(t_1)  =  \bm{v}_1, \\
&m(t_1) = m_i \\
&\bm{r}(t_2)  =  \bm{r_2}, \\
&\dot{\bm{r}}(t_2)  =  \bm{v}_2, \\
&\left|\bm{u}(t)\right|  \leqslant  T_{max} \quad \forall t\in [t_1, t_2], \\
&m_f \geqslant 0.
\end{array}
\right.
\end{equation}
where $t_2 = t_1 + tof$ is the arrival epoch, and $\mathcal F$ is some functional space (e.g., the space of all piecewise continuous functions defined in $[\underline t, \overline t + \overline{tof}]$). Note that the solution to this OCP is also minimizing the propellant mass $m_p = m_i - m_f$ used for the transfer.

To solve the OCP we use a direct transcription method \cite{yam} based on the Sims-Flanagan method, a well consolidated methodology for preliminary trajectory design \cite{sims}.

The method transforms the OCP into a non linear programming problem (NLP). The trajectory is divided into $n_{seg}$ segments of equal duration $\frac{tof}{n_{seg}}$ (we use $n_{seg}=100$). Along each segment a constant thrust vector is applied, computed by multiplying $T_{max}$ by a throttle vector represented by its Cartesian components $\bm{T}_i = \left\{T_i^x, T_i^y, T_i^z\right\}$, where $\lvert\bm{T}_i\rvert \leq 1$.

The dynamics is propagated along $n_{fwd} = \ceil{\frac{n_{seg}}{2}}$ segments (where $\ceil{\frac{n_{seg}}{2}}$ is the least integer that is greater than or equal to $\frac{n_{seg}}{2}$) to a mid-point $\bm{y}_{mf} = \left\{r_x, r_y, r_z, v_x, v_y, v_z, m\right\}_{mf}$ starting from the initial conditions and using the corresponding throttles. Starting from the final conditions, a second mid-point $\bm{y}_{mb}$ is computed propagating the dynamics backward along $n_{bck} = \floor{\frac{n_{seg}}{2}}$ segments (where $\floor{\frac{n_{seg}}{2}}$ is the greatest integer that is lesser than or equal to $\frac{n_{seg}}{2}$) and using the corresponding throttles.

The forward and backward-propagated half-trajectories meet at the matchpoint, i.e., the mismatch in position, velocity, and mass:
$$ \bm{y}_{mf} - \bm{y}_{mb}  = \left\{\Delta r_x, \Delta r_y, \Delta r_z, \Delta v_x, \Delta v_y, \Delta v_z, \Delta m\right\} = 0 $$
is added as a constraint to the resulting NLP.

\begin{figure}[t]
    \center
    \begin{minipage}{.49\textwidth}
        \centerline{\includegraphics[width=\textwidth]{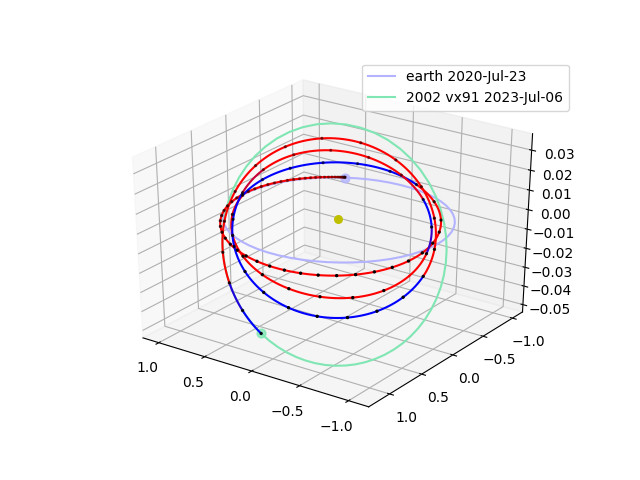}}
    \end{minipage}
    \begin{minipage}{.49\textwidth}
        \centerline{\includegraphics[width=\textwidth]{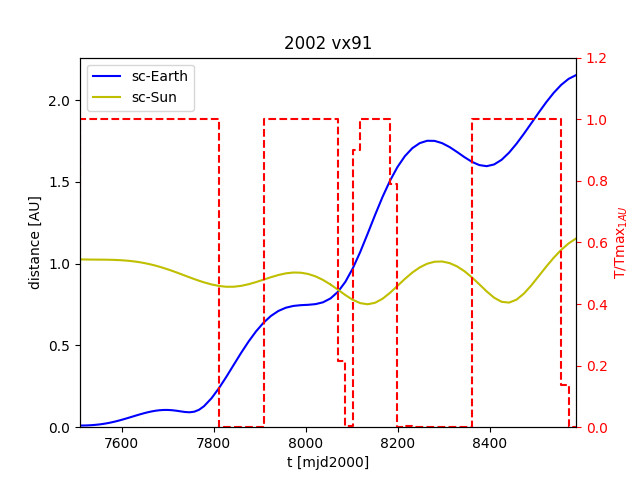}}
    \end{minipage}
    \caption{Example of a 3-year optimal trajectory using the NEP model.
    Left: the trajectory, with thrust arcs in \emph{red} and coast arcs
    in \emph{blue} (the coordinates on the axes are in AU).
    Right: thrust profile (\emph{red,
    dashed}) and distances of the spacecraft from the Sun
    (\emph{yellow}) and from the Earth (\emph{blue}) over time.}
    \label{fig:traj_NEP}
\end{figure}

More formally, introducing the decision vector:
$$
\bm{x} = \left\{ t_1, tof, m_f, T_1^x, T_1^y, T_1^z, \dotsc, T_{n_{seg}}^x, T_{n_{seg}}^y, T_{n_{seg}}^z \right\}
$$
the resulting NLP is described by:
\begin{equation}
\label{eq:ocp-nlp}
\mathcal P_3: \left\{
\begin{array}{rl}
\text{find:} & \bm{x} \in [\underline{\bm{x}}, \overline{\bm{x}}]\\
\text{to maximize:} & m_f \\
\text{subject to: } & \\
& \lvert\bm{T}_i\rvert \leq 1, \; i = 1, \ldots, n_{seg},\\
& \bm{y}_{mf} - \bm{y}_{mb} = 0.
\end{array}
\right.
\end{equation}
where the lower and upper bounds of the decision vector are given by, respectively:
$$
\underline{\bm{x}} = \left\{ \underline t, \underline{tof}, 0, -1, -1, -1, \dotsc, -1, -1, -1 \right\}
$$
and
$$
\overline{\bm{x}} = \left\{ \overline t, \overline{tof}, m_i, 1, 1, 1, \dotsc, 1, 1, 1 \right\}
$$
To solve the resulting NLP we use the software package called SNOPT \cite{snopt}, which implements a sequential quadratic programming (SQP) approach. We do not provide analytical derivatives and leave the software to numerically estimate them. As an initial guess we generate a random decision vector within the allowed bounds, and to propagate the equations of motion we use a Taylor integrator. The solutions found yield discrete values for the throttles arranged along a typical \emph{bang-bang} control profile as shown in Fig.(\ref{fig:traj_NEP}) and theoretically predicted by applying Pontryagin maximum principle to the stated OCP \cite{pont}. Note that the computed thrust profile shows some numerical imperfections, essentially due to two factors: numerical artifacts of the SQP solver, and the discretization of the time into segments. As known from previous studies on the Sims-Flanagan transcription \cite{sims, sepmalto} despite these effects it still result in a very good approximation of the final optimal propellant mass computed.

\subsection{NEP+G}

Adding the pull due to the Earth's gravity to the spacecraft dynamics
used in the NEP model, we obtain what is here indicated as NEP+G
model. The equations of motion Eq.(\ref{eqn:mot}) thus become:
\begin{eqnarray}
\label{eqn:mot2}
 \ddot{\bm{r}} &=& -\mu\frac{\bm{r}}{r_S^3} + \frac{\bm{u}(t)}{m} -
 \mu_E\frac{\bm{r_E}}{r_E^3}, \\
\dot{m} & = & -\frac{\left|\bm{u}(t)\right|}{I_{sp}g_0}, \nonumber
\end{eqnarray}
where $\mu_E \approx 3.986\cdot10^{14}$ m$^3$/s$^2$ is the
gravitational parameter of the Earth. The addition of the Earth's
gravity modifies the resulting OCP, but only marginally so that we can still transcribe it into a NLP dividing the trajectory into segments. In the new transcription, the Earth-spacecraft vector
$\bm{r_E}$ is also (similarily to the thrust) assumed constant along each
segment and is computed at the beginning of each forward-segment and at
the end of each backward-segment. This change makes the resulting
propagated trajectory only an approximate solution to the original
equations of motion, since $\bm{r_E}$ is forced to not vary along any given segment. The effect of this approximation on the resulting NLP solution will be particularly severe when the Earth-spacecraft distance is small, which
happens at the beginning of the trajectory and may happen later on
according to the particular case considered. In order to increase the accuracy we thus introduce a non uniform time grid
having a larger density at departure. The grid points epochs are
defined by:
$$ 
t^i = \begin{cases}
tof \cdot \left(\frac{i}{n_{seg}}\right)^2 & \text{for }i = 1, \dotsc, \floor{\frac{n_{seg}}{2}}\\
tof \cdot \left(1.5\frac{i}{n_{seg}}-0.5\right) & \text{for }i  = \floor{\frac{n_{seg}}{2}}+1, \dotsc, n_{seg}
\end{cases}
$$

\begin{figure}[t]
    \center
    \includegraphics[width=.6\textwidth]{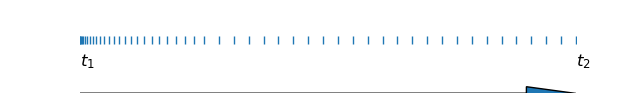}
    \caption{A visualization of the epochs grid used to capture the Earth's gravity effects accurately  ($n_{seg} = 50$).}
    \label{fig:timegrid}
\end{figure}

In Fig.(\ref{fig:timegrid}), we show a visualization of the time grid defined by the relations above in the case $n_{seg}=50$. Since the epochs are no longer evenly spaced, the propagation error will accumulate differently along each segment, according to the segment time length.  We thus take the mid point of the trajectory at half the time of flight (i.e., the first $t^i>\frac{tof}{2}$ marks the end of the last forward-segment and the beginning of the first backward-segment), which result in a number of forward-segments larger than the number of backward-segments. As an example, when $n_{seg}=100$ we have $n_{fwd} = 67$, $n_{bck}=33$.

\begin{figure}[t]
    \center
    \begin{minipage}{.49\textwidth}
        \centerline{\includegraphics[width=\textwidth]{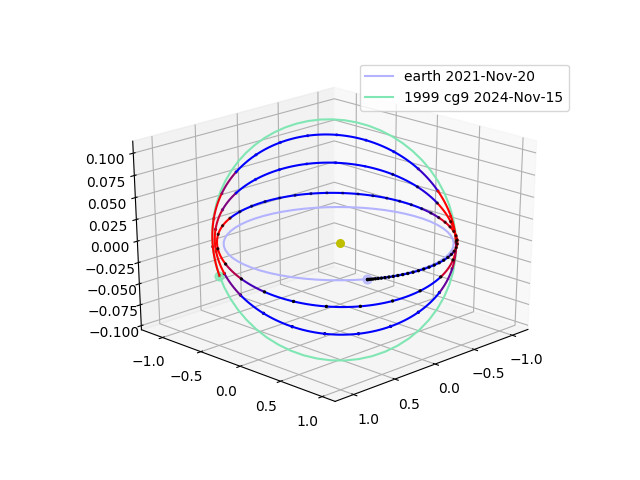}}
    \end{minipage}
    \begin{minipage}{.49\textwidth}
        \centerline{\includegraphics[width=\textwidth]{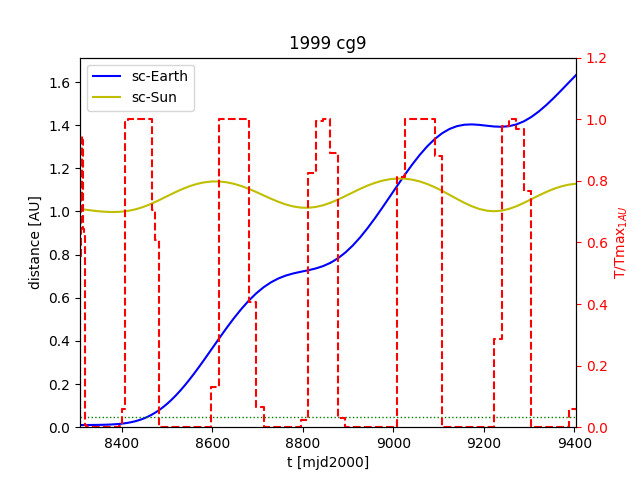}}
    \end{minipage}
    \caption{Example of a 3-year optimal trajectory using the NEP+G model. Left: the trajectory, with thrust arcs in \emph{red} and coast arcs
    in \emph{blue} (the coordinates on the axes are in AU).
    Right: thrust profile (\emph{red, dashed}) and distances of the spacecraft from the Sun (\emph{yellow}) and from the Earth (\emph{blue}) over time; the \emph{green, dashed} line delimits a sort of sphere of influence of the Earth where its gravity is greater than $0.1*T_{max}$.}
    \label{fig:traj_NEP+G}
\end{figure}

Moreover, some preliminary runs showed that the solution to the NLP, in some cases, produces trajectories attempting an Earth fly-by along the way, i.e., it places some grid-points very close to the Earth in order to take advantage of a large gravity vector for the following segment. When this happens, our model is not able to capture the real dynamics of the spacecraft, since it keeps $\bm{r_E}$ constant while it is instead rapidly changing when in proximity of the Earth. To avoid this behaviour we introduce an additional set of constraints in our NLP, preventing the Earth-spacecraft vector to become too small:
$$
\left|\bm{r_E}^i\right| \geq 0.01 AU, \; i = 1, \ldots, n_{seg}
$$
In Fig.(\ref{fig:traj_NEP+G}) we show an example of an optimal NEP+G trajectory found as solution to the NLP defined. The \emph{bang-bang} structure of the optimal thrust vector magnitude is well approximated also in this case.

\subsection{SEP+G}
\begin{figure}[t]
    \center
    \begin{minipage}{.49\textwidth}
        \centerline{\includegraphics[width=\textwidth]{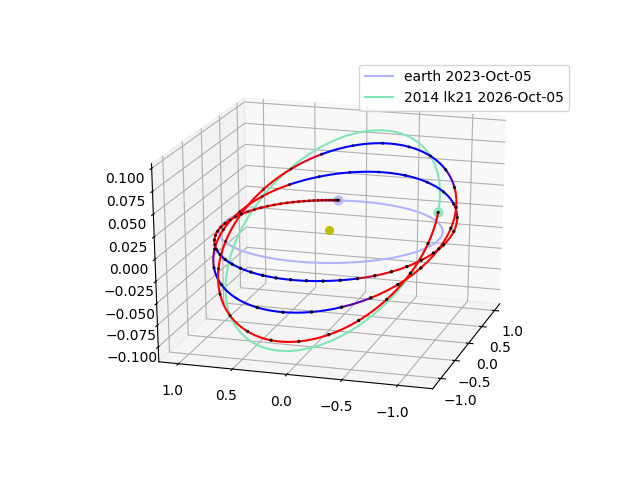}}
    \end{minipage}
    \begin{minipage}{.49\textwidth}
        \centerline{\includegraphics[width=\textwidth]{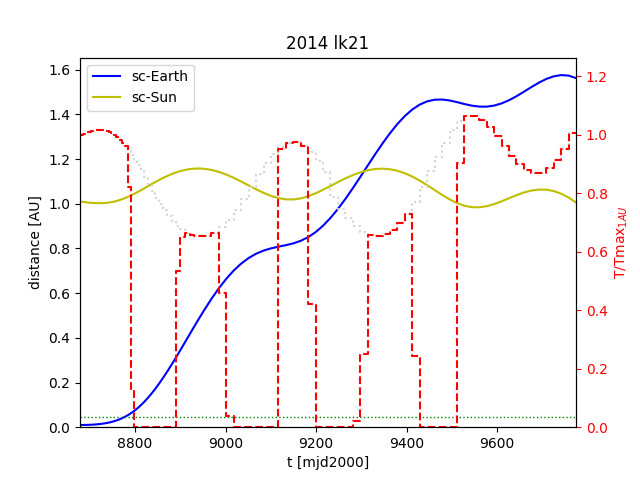}}
    \end{minipage}
    \caption{Example of a 3-year optimal trajectory using the SEP+G model. Left: the trajectory, with thrust arcs in \emph{red} and coast arcs
    in \emph{blue} (the coordinates on the axes are in AU).
    Right: thrust profile (\emph{red, dashed}) and distances of the spacecraft from the Sun (\emph{yellow}) and from the Earth (\emph{blue}) over time; note that $T_{max}$ (\emph{gray, dotted}) here depends on the Sun-spacecraft distance.}
    \label{fig:traj_SEP+G}
\end{figure}

Our last and most complete trajectory model is called SEP+G and adds to the NEP+G dynamics the effect of having a solar electric propulsion system. The spacecraft maximum thrust is thus no longer assumed to be constant, rather it is made dependent on the Sun-spacecraft distance according to the model introduced in Section \S \ref{sec:mission}. The distance from the Sun $\bm{r_S}$ is assumed constant along each segment and computed at the beginning of each forward-segment and at the end of each backward-segment along with the Earth's gravity and the spacecraft thrust vector. An example of an optimal SEP+G trajectory is reported in Fig.(\ref{fig:traj_SEP+G}).

\section{Asteroid database and pre-filtering}
\label{sec:prefilter}
\noindent Before optimizing and ranking the full trajectories using our increasingly complex trajectory models, we have to select a number of candidate targets we want to focus upon. Consideration of all the known asteroids under all our trajectory models would be computationally intractable as well as unnecessary since most asteroids are anyway not within reach of a small interplanetary CubeSat mission such as the M-ARGO spacecraft here considered. We thus perform a quick and conservative scan of the full MPCORB catalog of minor planets \cite{mpcorb} as of 7 June 2017 and we compute for each of them a $\Delta V_{3I}$ required for the transfer using the simplest of our models: the three impulses approximation (see Section \S \ref{ssec:imp}). Given the computational simplicity of such an approximation we are able to compute it for the full catalogue in a few seconds.
Since our M-ARGO mission profile prescribes a propellant mass of 2.5 kg, we use the rocket equation \cite{tsiolkovsky}, to compute a corresponding $\Delta V_{\mathrm{M-ARGO}}$ of 3928 m/s:
$$
\Delta V_{\mathrm{M-ARGO}} = I_{sp}g_0\ln{\frac{m_i}{m_i-m_p}}
$$
and use this as a filtering criteria $ \Delta V_{3I} \le \Delta V_{\mathrm{M-ARGO}} $.  In addition to that, we take only asteroids with an absolute magnitude $H \leq 26$ (such that their diameter is at least few dozens of meters) and we require a number of observations $n_{\mathrm{obs}} \geq 40$ to ensure a reasonable accuracy on their orbit. The result is a list of 143 asteroids to focus upon. Given the nature of the three impulses approximation such a list is likely to be over optimistic as the real $\Delta V$ necessary to reach each of them is likely to be significantly larger as it will have to account for phasing, for the argument of perihelion matching and for the gravity losses introduced by the low-thrust system. As we will show in the next section, this is indeed the case which indicates we have not pre-filtered out possible targets.

\section{Results}
\label{sec:results}
\subsection{Non-dominated fronts}
\noindent The OCP defined in Eq.(\ref{eq:OCP}) considers the final spacecraft mass as only objective for maximization. In reality the trajectory transfer time $tof = t_2-t_1$ also plays an important role in the preliminary mission design and many would argue that trajectory optimization is, in general, a multiobjective problem. As a consequence, when ranking different target candidates in terms of the quality of transfer opportunities one could compare the corresponding Pareto fronts using, for example, the hypervolume indicator as proposed in \cite{hv}. In the case of M-ARGO, given the simple nature of the transfer, and the $tof$ constraints, this is not considered as necessary and the final target selection and ranking is thus made considering a fixed $tof=3$ years. It is nevertheless interesting to compute and study the different non-dominated fronts. 

To compute, for a given target, a set of non-dominated solutions, we consider several bins for the $tof$ covering the 1-4 year range. We then solve the OCP bounding the transfer time in each bin and using the NEP, NEP+G and SEP+G models. The obtained set of non-dominated solutions for each model will approximate the corresponding Pareto front. We were thus able to compute the set of non-dominated solutions for all the 143 asteroid considered.

\begin{figure}[t]
    \center
    \begin{minipage}{.49\textwidth}
        \centerline{\includegraphics[width=\textwidth]{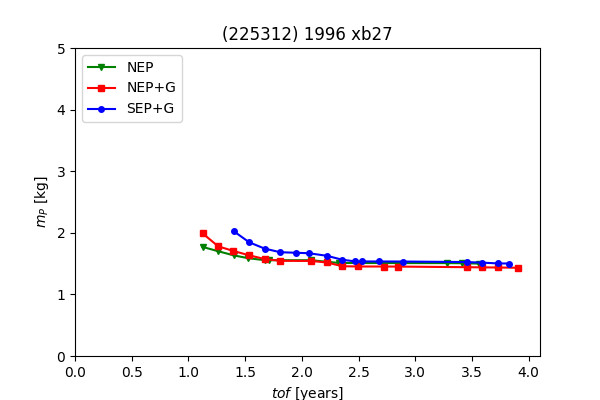}}
        \caption{Pareto fronts for the asteroid \emph{(225312) 1996 XB$_{27}$}.}
        \label{fig:pareto1}
    \end{minipage}
    \begin{minipage}{.49\textwidth}
        \centerline{\includegraphics[width=\textwidth]{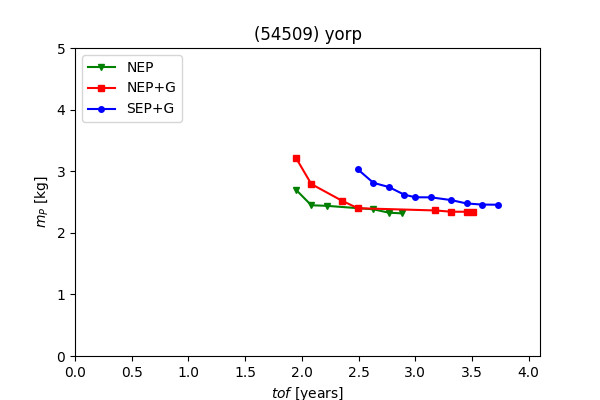}}
        \caption{Pareto fronts for the asteroid \emph{(54509) YORP}.}
        \label{fig:pareto2}
    \end{minipage}
\end{figure}

As examples, we briefly discuss some of the non dominated fronts computed in selected cases. In Fig.(\ref{fig:pareto1}) the case for asteroid (225312) 1996 XB$_{27}$ is considered. The asteroid orbit is characterized by a very low inclination and eccentricity and its semi major axis is $a=1.19AU$, making it a rather easy target to reach in the considered timeframe. This reflects immediately on the set of non-dominated solutions computed that appears rather flat with an propellant consumption consistently small ($m_p < 2$ kg) across transfer times. Note that the use of different trajectory models, in this case, does not change significantly the set of non dominated solutions. Only in the SEP+G case we note a slight shift towards a higher propellant cost as obvious since less thrust will be available as the spacecraft spirals out toward this target.

\begin{figure}[t]
    \center
    \begin{minipage}{.49\textwidth}
        \centerline{\includegraphics[width=\textwidth]{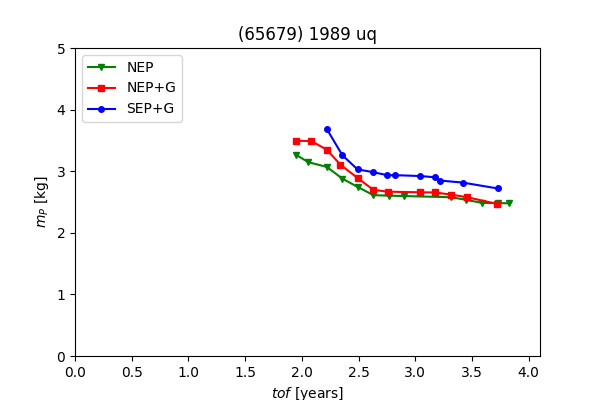}}
        \caption{Pareto fronts for the asteroid \emph{(65679) 1989 UQ}.}
        \label{fig:pareto3}
    \end{minipage}
    \begin{minipage}{.49\textwidth}
        \centerline{\includegraphics[width=\textwidth]{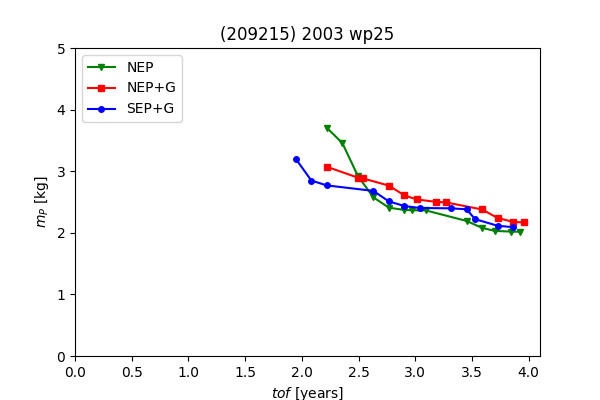}}
        \caption{Pareto fronts for the asteroid \emph{(209215) 2003 WP$_{25}$}.}
        \label{fig:pareto4}
    \end{minipage}
\end{figure}

In Fig.(\ref{fig:pareto2}) the case for asteroid (54509) YORP is considered. This asteroid, famous for being the first to allow the YORP effect measurement, has a slight inclination $i=1.83^{\circ}$, the same orbital period as the Earth and an eccentricity of $e=0.23$ which makes it an Earth-crossing asteroid. The close proximity of the orbital period to the Earth's makes phasing a greater concern. Such a different orbital configuration is also reflected directly on the set of non-dominated solutions found that are now favouring considerably longer transfer times (allowing for a more efficient phase correction) and are also distinguishing more clearly between the three low-thrust trajectory models. A similar non dominated set of optimal solutions is found - and shown in Fig.(\ref{fig:pareto3}) - for the asteroid (65679) 1989 UQ, having similar characteristics. Note that a consistently increased propellant consumption of the SEP+G model is not always the case and, in asteroids getting closer to the Sun, one may also find that the SEP+G model results in a more efficient use of the propellant, as is the case of asteroid (209215) 2003 WP$_{25}$ reported in  Fig.(\ref{fig:pareto4}).

All in all, the set of non-dominated solutions, while not commonly computed, offers an interesting chart to be produced at preliminary stages as it helps quantifying early on the trade-off between propellant mass and transfer time, providing system engineers with one important piece of information to tune their baseline design around.

\subsection{Target selection and ranking}
\noindent In order to rank the 143 pre-selected targets we have to choose a ranking criteria. While the hypervolume of the non-dominated fronts is an interesting metric to consider, we also know that, for the mission profile considered, the longer the $tof$ the smaller the $m_p$ will be. Since we are interested in trajectories with $tof < 3$ years,  we take as a ranking criterion the $m_p$ required when we fix $tof=3$ years. This will also allow us to compare the ranks obtained to those computed using the impulsive models for which the set of non-dominated solutions cannot be computed.

Our ground truth is given by the SEP+G model, it being the most accurate. But accuracy comes at the cost of significantly longer computation times, hence our interest in measuring the correlations between the rankings resulting from the five trajectory models. We compute these correlations using the Kendall's tau coefficient: given two different rankings of the asteroids $R_i = \left[ \mathcal{A}_{i_1}, \dotsc,\mathcal{A}_{i_n} \right]$ and $R_j = \left[ \mathcal{A}_{j_1}, \dotsc,\mathcal{A}_{j_n} \right]$, it is defined as:
\begin{equation}
    \tau = \frac{n_c-n_d}{n(n-1)/2}
\end{equation}
where $n_c$ is the number of concordant pairs and $n_d$ is the number of discordant pairs in the two rankings. A value of $\tau = 1$ corresponds to two identical rankings, while $\tau = -1$ corresponds to two perfectly discordant rankings and $\tau = 0$ represents absence of correlation. We can also transform the Kendall tau coefficient into the percentage of wrongly classified pairs if one of the two rankings, as is our case, is to be considered as the ground truth: 
$$
err = \frac{n_d}{n(n-1)/2} = \frac{1-\tau}{2}
$$
so that, for example, a $\tau = 0.5$ corresponds to $err=0.25$, that is to 25\% of incorrectly ranked pairs. 

Using the five trajectory models described in Section \S \ref{sec:traj_models} we compute $m_p$ for all of the 143 asteroids and assemble the five corresponding ranks $R_{3I}, R_{L}, R_{\mathrm{NEP}},R_{\mathrm{NEP+G}},R_{\mathrm{SEP+G}}$. We are then able to compute the rank correlation metric $\tau$ for any two rankings. 

\begin{table}[t]
    \begin{center}
    \caption{Kendall tau coefficient for each pair of rankings. The greater the value, the higher the rank correlation. The bold values correspond to the comparison with our SEP+G model considered as the ground truth}
    \label{tab:ktau}
    \vspace {-3.mm}
    \begin{tabular}{ l | l | l | l | l | l }
    \toprule
     & 3I & L & NEP & NEP+G & SEP+G \\[.5mm]
    \hline
    3I & - & 0.577 & 0.441 & 0.421 & \bf{0.432} \\
    L & 0.577 & - & 0.660 & 0.646 & \bf{0.628} \\
    NEP &  0.441 & 0.660 & - & 0.864 & \bf{0.841} \\
    NEP+G & 0.421  & 0.646 & 0.864 & - & \bf{0.845} \\
    SEP+G & \bf{0.432} & \bf{0.628} & \bf{0.841} & \bf{0.845} & \bf{-}  \\
    \bottomrule
    \end{tabular}
    \end{center}
\end{table}

The results are contained in Table \ref{tab:ktau}. Since SEP+G is our reference model (the ground truth) the last column of the table is the most interesting and significant. Its shows quantitatively the error we introduce when performing target selection using a simplified dynamical model. While the table is certainly accurate for the study case here considered (M-ARGO) it is also useful as an indicator of a more general trend. The simple three impulses approximation, while very fast and efficient, has the lowest correlation to the ground truth with $\tau = 0.432$, corresponding to a 28.5\% probability of mis-ranking any given pair of asteroids. The use of a Lambert model improves things considerably adding the extra cost of having to solve the optimization problem stated in Eq.(\ref{eq:lambertP}) for each of the 143 targets. The rank correlation coefficient is $\tau = 0.628$ corresponding to a 18.6\% probability of mis-ranking asteroid pairs. The use of the NEP model introduces a net improvement as the rank correlation becomes $\tau=0.841$ and thus the probability to misrank asteroid pairs becomes only 7.95\%. The NLP defined in Eq.(\ref{eq:ocp-nlp}) needs to be solved for each of the 143 targets, which adds complexity with respect to the Lambert case, but also increases the resulting rank precision considerably. The last model considered (excluding the ground truth) is the NEP+G model that, essentially, creates a ranking that correlates only slightly better to the ground truth than the NEP model, having $\tau = 0.845$ and 7.75\% probability to misrank asteroid pairs. Such a small improvement seems to hardly justify the inclusion of Earth's gravity into the dynamical model since this also adds complexity to the resulting OCP/NLP.

The Kendall tau rank correlation is a great quantification of the effects introduced by the choice of the model on the final target selection, but sometimes we are only interested in using the computed $m_p$ to simply classify some target as reachable or not. In our study case, the criteria for a certain target to be classified as reachable by our spacecraft would be $m_p\le2.5$ kg as that is dictated by the system design of the M-ARGO tanks. Assuming once again the SEP+G model as our ground truth, we can then compute the confusion matrix if such a classification were made using propellant masses computed using each of the approximated models. The results are shown in Table \ref{tab:conf_lambert}. Out of the 85 asteroids that are actually reachable by M-ARGO (according to the SEP+G optimized trajectories) the NEP and NEP+G models do not allow to select, respectively, 2 and 3 possible targets, while they wrongly select 13 and 9 as reachable when they are not. The simpler Lambert model, instead, misses 15 potentially interesting targets and wrongly locates 15 asteroids as reachable.

\begin{table}[t]
    \center
    \caption{Confusion matrices for the classification of the asetroids as reachable targets.}
    \begin{center}
    \vspace {-3.mm}
    \begin{tabular}{c c|c|c}
        \toprule
        \label{tab:conf_lambert}
        & & \multicolumn{2}{c}{$m_p^{\mathrm{SEP+G}}$}\\
        & & $\leq2.5$ & $>2.5$ \\
        \hline
        \multicolumn{1}{c}{\multirow{2}{*}{$m_p^{\mathrm{L}}$}} & $\leq2.5$ & 70 & 15 \\
        \cline{2-4}
        \multicolumn{1}{c}{} & $>2.5$ & 15 & 43 \\
        \bottomrule
    \end{tabular}
    \quad
    \begin{tabular}{c c|c|c}
        \toprule
        \label{tab:conf_NEP}
        & & \multicolumn{2}{c}{$m_p^{\mathrm{SEP+G}}$}\\
        & & $\leq2.5$ & $>2.5$ \\
        \hline
        \multicolumn{1}{c}{\multirow{2}{*}{$m_p^{\mathrm{NEP}}$}} & $\leq2.5$ & 83 & 13 \\
        \cline{2-4}
        \multicolumn{1}{c}{} & $>2.5$ & 2 & 45 \\
        \bottomrule
    \end{tabular}
    \quad
    \begin{tabular}{c c|c|c}
        \toprule
        \label{tab:conf_NEP+G}
        & & \multicolumn{2}{c}{$m_p^{\mathrm{SEP+G}}$}\\
        & & $\leq2.5$ & $>2.5$ \\
        \hline
        \multicolumn{1}{c}{\multirow{2}{*}{$m_p^{\mathrm{NEP+G}}$}} & $\leq2.5$ & 82 & 9 \\
        \cline{2-4}
        \multicolumn{1}{c}{} & $>2.5$ & 3 & 49 \\
        \bottomrule
    \end{tabular}
    \qquad
    \end{center}
\end{table}

Finally, in Table (\ref{tab:ranking}), we report the list of asteroids ranked (using the $m_p$ computed by the SEP+G model) for our M-ARGO study case. The final propellant mass, as computed with the five trajectory models, is reported, as well as the trajectory launch date and data on the target asteroid. 

\begingroup
\renewcommand\arraystretch{0.7}
\setlength{\tabcolsep}{5pt}
\begin{longtable}{ l | l | l | l | l | l | l | l | r | r | r | r }
    \caption{Final list of possible targets, ranked according to the SEP+G model. $m_p$ is the propellant mass required for an optimal 3-year transfer using the five models; $t_1$ is the launch date; $H$ is the absolute magnitude; $d$ is the diameter estimated from $H$; $n_{\mathrm{obs}}$ is the number of observations; $n_{\mathrm{opp}}$ is the number of oppositions.\label{tab:ranking}} \\

    \toprule
    rank & name & $m_p^{\mathrm{3I}}$ & $m_p^{\mathrm{L}}$ & $m_p^{\mathrm{NEP}}$ & $m_p^{\mathrm{NEP+G}}$ & $m_p^{\mathrm{SEP+G}}$ & $t_1$ & $H$ & $d$ & $n_{\mathrm{obs}}$ & $n_{\mathrm{opp}}$ \\
    &  & [kg] & [kg] & [kg] & [kg] & [kg] &  &  & [m] &  &  \\[.5mm]
    \hline
    \endfirsthead
    \hline
    rank & name & $m_p^{\mathrm{3I}}$ & $m_p^{\mathrm{L}}$ & $m_p^{\mathrm{NEP}}$ & $m_p^{\mathrm{NEP+G}}$ & $m_p^{\mathrm{SEP+G}}$ & $t_1$ & $H$ & $d$ & $n_{\mathrm{obs}}$ & $n_{\mathrm{opp}}$ \\
    &  & [kg] & [kg] & [kg] & [kg] & [kg] &  &  & [m] &  &  \\[.5mm]
    \hline
    \endhead
    1 & 2016 tb57 & 1.06 & 1.18 & 0.89 & 0.94 & 0.79 & 2023-Jul-28 & 26.0 & 16-37 & 135 & 1 \\
    2 & 2013 wa44 & 1.40 & 1.21 & 1.05 & 1.10 & 1.07 & 2020-Apr-03 & 23.7 & 48-108 & 65 & 1 \\
    3 & 2013 bs45 & 0.74 & 1.10 & 1.21 & 1.36 & 1.16 & 2021-Apr-24 & 25.9 & 17-39 & 92 & 2 \\
    4 & 2016 cf137 & 1.51 & 1.47 & 1.21 & 1.19 & 1.19 & 2023-Jul-22 & 25.6 & 20-45 & 50 & 1 \\
    5 & 2014 yd & 1.18 & 1.18 & 1.42 & 1.23 & 1.22 & 2023-Apr-13 & 24.3 & 36-82 & 104 & 1 \\
    6 & 2015 bm510 & 1.18 & 1.54 & 1.26 & 1.34 & 1.32 & 2023-Apr-18 & 25.1 & 25-56 & 58 & 1 \\
    7 & 2014 sd304 & 1.81 & 1.71 & 1.35 & 1.38 & 1.35 & 2023-Aug-05 & 25.0 & 26-59 & 63 & 2 \\
    8 & 2012 ec & 1.45 & 1.80 & 1.40 & 1.36 & 1.35 & 2021-Apr-03 & 23.4 & 55-124 & 139 & 2 \\
    9 & 2009 cv & 1.41 & 1.57 & 1.28 & 1.37 & 1.42 & 2022-Apr-04 & 24.3 & 36-82 & 174 & 4 \\
    10 & 2009 os5 & 1.53 & 1.57 & 1.39 & 1.43 & 1.47 & 2020-Apr-03 & 24.5 & 33-74 & 68 & 2 \\
    11 & 2004 jn1 & 1.66 & 1.66 & 1.44 & 1.47 & 1.49 & 2020-Oct-04 & 23.6 & 50-113 & 82 & 2 \\
    12 & 2003 sm84 & 1.70 & 1.72 & 1.46 & 1.50 & 1.50 & 2021-May-03 & 22.7 & 76-171 & 98 & 3 \\
    13 & (478784) 2012 uv136 & 1.47 & 1.80 & 1.51 & 1.50 & 1.53 & 2023-Oct-05 & 25.5 & 21-47 & 121 & 5 \\
    14 & (225312) 1996 xb27 & 1.91 & 1.81 & 1.53 & 1.46 & 1.53 & 2022-Apr-04 & 21.7 & 121-271 & 170 & 5 \\
    15 & 2017 bf29 & 1.99 & 1.79 & 1.60 & 1.54 & 1.55 & 2020-Sep-30 & 25.6 & 20-45 & 45 & 1 \\
    16 & 2001 qj142 & 1.55 & 1.83 & 1.54 & 1.57 & 1.56 & 2022-May-22 & 23.7 & 48-108 & 91 & 2 \\
    17 & 2013 em89 & 1.90 & 1.69 & 1.75 & 1.51 & 1.56 & 2021-Apr-03 & 26.0 & 16-37 & 58 & 1 \\
    18 & 2012 hk31 & 1.47 & 1.91 & 1.60 & 1.66 & 1.61 & 2021-May-03 & 25.4 & 22-49 & 63 & 1 \\
    19 & 2013 pa7 & 2.01 & 1.85 & 1.59 & 1.57 & 1.61 & 2022-Aug-12 & 22.6 & 80-179 & 89 & 1 \\
    20 & 2015 pl57 & 1.56 & 1.86 & 1.57 & 1.67 & 1.65 & 2020-Apr-03 & 25.7 & 19-43 & 46 & 1 \\
    21 & 2016 tb18 & 1.13 & 1.19 & 1.54 & 1.58 & 1.69 & 2023-Apr-28 & 24.8 & 29-65 & 92 & 1 \\
    22 & 2012 uw68 & 1.86 & 2.04 & 1.76 & 1.78 & 1.70 & 2023-Oct-05 & 24.8 & 29-65 & 43 & 1 \\
    23 & 2017 hk1 & 1.40 & 1.80 & 1.65 & 1.59 & 1.73 & 2021-Sep-11 & 25.0 & 26-59 & 77 & 1 \\
    24 & 2004 vj1 & 1.51 & 1.85 & 1.65 & 1.67 & 1.73 & 2023-Apr-04 & 24.2 & 38-85 & 97 & 2 \\
    25 & 2015 tz24 & 2.16 & 1.99 & 1.78 & 1.69 & 1.74 & 2023-Apr-07 & 24.1 & 40-89 & 42 & 1 \\
    26 & 1999 ao10 & 1.31 & 1.82 & 1.71 & 1.77 & 1.75 & 2023-Apr-04 & 23.9 & 44-98 & 73 & 1 \\
    27 & 2014 yn & 1.30 & 1.79 & 1.71 & 1.79 & 1.76 & 2023-Apr-06 & 25.7 & 19-43 & 78 & 1 \\
    28 & 2007 tf15 & 1.95 & 1.87 & 1.68 & 1.70 & 1.78 & 2021-Apr-03 & 24.6 & 31-71 & 60 & 2 \\
    29 & 2001 cq36 & 1.63 & 1.87 & 1.75 & 1.97 & 1.80 & 2020-Jul-21 & 22.5 & 84-187 & 133 & 4 \\
    30 & (459872) 2014 ek24 & 1.81 & 1.88 & 1.77 & 1.85 & 1.81 & 2023-Jul-12 & 23.3 & 58-130 & 572 & 2 \\
    31 & 2010 ha & 1.89 & 1.95 & 1.72 & 1.76 & 1.85 & 2023-Sep-27 & 23.9 & 44-98 & 62 & 2 \\
    32 & 2005 tg50 & 1.42 & 1.76 & 1.70 & 1.84 & 1.85 & 2020-Apr-03 & 24.8 & 29-65 & 70 & 2 \\
    33 & 1999 cg9 & 2.07 & 1.88 & 1.81 & 1.88 & 1.90 & 2021-Oct-03 & 25.2 & 24-54 & 42 & 1 \\
    34 & 2008 tx3 & 2.08 & 2.24 & 1.91 & 1.92 & 1.93 & 2021-Aug-26 & 24.9 & 27-62 & 101 & 1 \\
    35 & 2006 fh36 & 1.84 & 2.04 & 1.86 & 1.98 & 1.96 & 2020-Oct-03 & 22.9 & 69-156 & 62 & 2 \\
    36 & 2007 uy1 & 1.59 & 2.28 & 1.90 & 1.97 & 1.97 & 2020-May-03 & 22.9 & 69-156 & 130 & 2 \\
    37 & 2013 xy20 & 1.79 & 1.81 & 1.59 & 1.92 & 1.98 & 2023-Sep-24 & 25.5 & 21-47 & 59 & 1 \\
    38 & 2015 vv & 2.12 & 1.93 & 1.71 & 1.68 & 1.99 & 2021-May-01 & 24.3 & 36-82 & 183 & 1 \\
    39 & 2006 qv89 & 2.02 & 2.33 & 2.12 & 2.04 & 2.03 & 2022-Jul-16 & 25.3 & 23-51 & 68 & 1 \\
    40 & 2016 vl3 & 2.41 & 2.47 & 2.02 & 1.95 & 2.05 & 2021-Aug-28 & 24.4 & 35-78 & 62 & 1 \\
    41 & 1998 ky26 & 2.07 & 2.25 & 2.10 & 2.12 & 2.07 & 2023-Apr-04 & 25.5 & 21-47 & 211 & 1 \\
    42 & 2000 ae205 & 2.40 & 2.22 & 2.06 & 2.02 & 2.08 & 2023-Jul-16 & 23.0 & 66-149 & 80 & 3 \\
    43 & 2007 dd & 1.39 & 1.81 & 2.14 & 2.38 & 2.08 & 2022-Oct-04 & 25.8 & 18-41 & 76 & 4 \\
    44 & 2014 mf18 & 1.69 & 1.75 & 1.99 & 2.07 & 2.08 & 2022-Oct-03 & 26.0 & 16-37 & 55 & 1 \\
    45 & 2009 rt1 & 2.23 & 2.16 & 2.01 & 2.05 & 2.09 & 2023-Apr-08 & 23.6 & 50-113 & 45 & 1 \\
    46 & 2015 lj & 2.14 & 2.16 & 1.97 & 2.01 & 2.11 & 2022-Apr-06 & 24.7 & 30-68 & 107 & 1 \\
    47 & (251732) 1998 hg49 & 2.42 & 2.22 & 2.17 & 2.10 & 2.12 & 2022-Jun-08 & 21.7 & 121-271 & 109 & 5 \\
    48 & 2016 fy2 & 1.73 & 2.17 & 2.16 & 2.03 & 2.12 & 2022-Oct-03 & 25.5 & 21-47 & 130 & 1 \\
    49 & 2003 ln6 & 1.96 & 2.07 & 2.25 & 2.27 & 2.12 & 2020-Sep-21 & 24.6 & 31-71 & 97 & 2 \\
    50 & 2013 hp11 & 2.37 & 2.20 & 2.01 & 2.05 & 2.12 & 2021-Apr-03 & 25.4 & 22-49 & 84 & 1 \\
    51 & 2014 uy & 2.30 & 2.34 & 2.14 & 2.11 & 2.13 & 2022-Sep-24 & 25.4 & 22-49 & 100 & 1 \\
    52 & 2017 eb3 & 1.74 & 2.32 & 2.05 & 2.13 & 2.13 & 2020-Apr-03 & 24.9 & 27-62 & 41 & 1 \\
    53 & (163000) 2001 sw169 & 2.43 & 2.69 & 2.05 & 2.05 & 2.15 & 2021-Apr-04 & 19.2 & 384-859 & 769 & 6 \\
    54 & 2009 hc & 1.84 & 2.14 & 2.01 & 2.10 & 2.18 & 2023-Apr-04 & 24.7 & 30-68 & 145 & 1 \\
    55 & 2015 tj1 & 2.29 & 2.26 & 2.23 & 2.09 & 2.23 & 2022-Aug-26 & 22.6 & 80-179 & 257 & 1 \\
    56 & 2005 er95 & 2.37 & 2.12 & 2.08 & 2.14 & 2.24 & 2023-Apr-04 & 25.4 & 22-49 & 47 & 1 \\
    57 & 2014 qh33 & 1.99 & 2.60 & 2.31 & 2.26 & 2.25 & 2021-Apr-06 & 24.4 & 35-78 & 108 & 2 \\
    58 & 2001 qe71 & 1.89 & 2.42 & 2.15 & 2.17 & 2.25 & 2020-Apr-04 & 24.4 & 35-78 & 68 & 1 \\
    59 & 2011 aa37 & 1.71 & 2.62 & 2.20 & 2.37 & 2.28 & 2023-Oct-05 & 22.8 & 73-163 & 54 & 1 \\
    60 & 2017 hz4 & 2.03 & 2.51 & 2.25 & 2.27 & 2.28 & 2021-Jun-28 & 25.8 & 18-41 & 46 & 1 \\
    61 & 2014 fa44 & 2.21 & 2.17 & 2.04 & 2.13 & 2.28 & 2022-Apr-04 & 24.8 & 29-65 & 105 & 1 \\
    62 & 2001 bb16 & 1.83 & 2.09 & 2.25 & 2.29 & 2.31 & 2021-Apr-05 & 23.2 & 60-136 & 45 & 3 \\
    63 & 2012 uy68 & 2.37 & 2.54 & 2.32 & 2.27 & 2.32 & 2020-Sep-10 & 25.0 & 26-59 & 46 & 1 \\
    64 & 2001 av43 & 2.07 & 2.31 & 2.32 & 2.18 & 2.32 & 2022-Jul-27 & 24.6 & 31-71 & 101 & 2 \\
    65 & 2012 wh & 1.84 & 2.12 & 2.35 & 2.59 & 2.32 & 2023-Apr-05 & 25.5 & 21-47 & 43 & 1 \\
    66 & 2016 ue & 1.39 & 1.73 & 2.12 & 2.10 & 2.33 & 2020-Jun-27 & 25.1 & 25-56 & 69 & 1 \\
    67 & 2016 cf194 & 2.32 & 2.35 & 2.21 & 2.24 & 2.33 & 2021-Apr-03 & 24.1 & 40-89 & 136 & 1 \\
    68 & 2015 fg36 & 2.09 & 2.52 & 2.29 & 2.39 & 2.35 & 2022-Apr-04 & 23.7 & 48-108 & 76 & 2 \\
    69 & 2006 xp4 & 1.98 & 2.38 & 2.35 & 2.26 & 2.35 & 2022-Apr-24 & 23.9 & 44-98 & 145 & 2 \\
    70 & 2013 rv9 & 2.36 & 2.56 & 2.33 & 2.34 & 2.35 & 2022-Aug-29 & 23.6 & 50-113 & 88 & 2 \\
    71 & 2012 ba35 & 2.41 & 2.52 & 2.34 & 2.45 & 2.38 & 2021-Jul-05 & 23.8 & 46-103 & 48 & 1 \\
    72 & 2010 wr7 & 2.13 & 2.52 & 2.43 & 2.62 & 2.40 & 2023-Apr-07 & 23.5 & 53-118 & 56 & 2 \\
    73 & 2017 bf30 & 1.83 & 2.34 & 2.30 & 2.35 & 2.40 & 2020-Jul-18 & 24.1 & 40-89 & 46 & 1 \\
    74 & 2011 cg2 & 2.07 & 2.24 & 2.24 & 2.15 & 2.41 & 2023-Jun-03 & 21.4 & 139-311 & 369 & 2 \\
    75 & 2016 tp11 & 1.66 & 2.23 & 2.09 & 2.14 & 2.44 & 2020-Apr-25 & 24.3 & 36-82 & 82 & 1 \\
    76 & (209215) 2003 wp25 & 1.41 & 1.96 & 2.46 & 2.57 & 2.45 & 2020-Jun-17 & 24.2 & 38-85 & 64 & 6 \\
    77 & 2016 dj & 2.12 & 2.57 & 2.31 & 2.40 & 2.46 & 2020-Oct-01 & 25.6 & 20-45 & 72 & 1 \\
    78 & 2001 ve2 & 2.37 & 2.77 & 2.43 & 2.40 & 2.47 & 2022-Sep-16 & 25.0 & 26-59 & 64 & 1 \\
    79 & 2010 vz11 & 2.33 & 2.77 & 2.48 & 2.49 & 2.47 & 2023-May-29 & 25.6 & 20-45 & 72 & 1 \\
    80 & 2014 un114 & 1.81 & 2.45 & 2.95 & 2.40 & 2.47 & 2023-Apr-24 & 24.5 & 33-74 & 177 & 1 \\
    81 & 2012 ux136 & 2.38 & 2.74 & 2.42 & 2.45 & 2.48 & 2020-Oct-04 & 25.6 & 20-45 & 50 & 1 \\
    82 & 2015 yk & 2.20 & 2.30 & 2.05 & 2.01 & 2.48 & 2023-Jun-01 & 25.9 & 17-39 & 91 & 1 \\
    83 & 2015 xc352 & 2.07 & 2.62 & 2.31 & 2.35 & 2.49 & 2020-Apr-03 & 25.7 & 19-43 & 75 & 2 \\
    84 & 2008 fo & 2.37 & 2.65 & 2.60 & 2.49 & 2.50 & 2020-Oct-04 & 23.2 & 60-136 & 91 & 1 \\
    85 & 1999 sf10 & 2.32 & 2.44 & 2.42 & 2.40 & 2.50 & 2021-Jul-31 & 24.3 & 36-82 & 54 & 3 \\
    86 & 2014 mz17 & 2.06 & 2.19 & 2.05 & 2.17 & 2.51 & 2023-Oct-05 & 24.1 & 40-89 & 47 & 2 \\
    87 & 2012 xf55 & 2.37 & 2.62 & 2.48 & 3.22 & 2.51 & 2021-Oct-05 & 22.8 & 73-163 & 56 & 1 \\
    88 & 2009 yf & 1.17 & 1.56 & 2.25 & 2.51 & 2.51 & 2023-Jul-05 & 24.7 & 30-68 & 44 & 1 \\
    89 & 2012 pb20 & 2.32 & 2.50 & 2.47 & 2.48 & 2.52 & 2023-Apr-19 & 24.9 & 27-62 & 45 & 1 \\
    90 & 2017 by & 2.42 & 2.52 & 2.46 & 2.47 & 2.56 & 2022-Sep-26 & 25.5 & 21-47 & 57 & 1 \\
    91 & (54509) yorp & 2.10 & 2.79 & 2.33 & 2.45 & 2.58 & 2020-Apr-03 & 22.7 & 76-171 & 533 & 5 \\
    92 & 2004 xk3 & 2.35 & 2.88 & 2.57 & 2.63 & 2.58 & 2022-Jun-21 & 24.4 & 35-78 & 229 & 2 \\
    93 & (450237) 2002 xy38 & 2.07 & 2.76 & 2.50 & 2.50 & 2.61 & 2020-Apr-05 & 22.9 & 69-156 & 110 & 3 \\
    94 & 2017 bu & 2.26 & 2.58 & 2.45 & 2.55 & 2.61 & 2020-Apr-05 & 25.1 & 25-56 & 59 & 1 \\
    95 & 2015 xa379 & 2.26 & 2.11 & 2.10 & 2.09 & 2.62 & 2021-Aug-21 & 25.3 & 23-51 & 275 & 1 \\
    96 & 2006 bz147 & 1.02 & 1.84 & 2.67 & 3.01 & 2.63 & 2023-Jul-23 & 25.4 & 22-49 & 61 & 3 \\
    97 & 2010 ps66 & 2.46 & 2.52 & 2.53 & 2.56 & 2.66 & 2020-Apr-28 & 25.1 & 25-56 & 74 & 1 \\
    98 & 2006 ct & 2.27 & 2.85 & 2.58 & 2.60 & 2.66 & 2020-Jun-13 & 22.2 & 96-215 & 108 & 4 \\
    99 & 2012 dk4 & 2.09 & 2.68 & 2.60 & 2.63 & 2.67 & 2020-May-26 & 23.6 & 50-113 & 63 & 1 \\
    100 & (350751) 2002 aw & 2.23 & 2.90 & 2.45 & 2.51 & 2.68 & 2021-Apr-10 & 20.8 & 183-411 & 240 & 5 \\
    101 & 2009 bf2 & 2.35 & 2.87 & 2.60 & 2.63 & 2.69 & 2020-Jun-28 & 25.9 & 17-39 & 64 & 1 \\
    102 & 2005 cn & 1.83 & 2.34 & 2.25 & 2.39 & 2.71 & 2020-May-06 & 22.8 & 73-163 & 61 & 6 \\
    103 & (350523) 2000 ea14 & 2.30 & 2.44 & 2.57 & 2.83 & 2.74 & 2023-Oct-05 & 21.1 & 160-358 & 143 & 4 \\
    104 & 2012 uk171 & 2.09 & 2.51 & 2.64 & 2.70 & 2.74 & 2022-Apr-05 & 24.5 & 33-74 & 201 & 2 \\
    105 & 2017 fb3 & 1.66 & 1.90 & 2.31 & 2.22 & 2.75 & 2020-May-16 & 25.8 & 18-41 & 49 & 1 \\
    106 & 2001 km20 & 2.49 & 3.12 & 2.73 & 2.71 & 2.76 & 2022-Apr-26 & 23.6 & 50-113 & 67 & 1 \\
    107 & 2017 bw & 2.24 & 2.69 & 2.83 & 3.05 & 2.77 & 2023-Jun-14 & 23.4 & 55-124 & 431 & 1 \\
    108 & 2010 xf3 & 2.21 & 2.50 & 2.56 & 2.71 & 2.77 & 2022-Jul-11 & 24.4 & 35-78 & 94 & 1 \\
    109 & 2000 uk11 & 2.29 & 2.57 & 2.51 & 2.61 & 2.79 & 2023-Apr-05 & 25.3 & 23-51 & 66 & 2 \\
    110 & 2011 gr59 & 2.32 & 2.80 & 2.74 & 2.75 & 2.79 & 2020-Apr-03 & 23.7 & 48-108 & 43 & 1 \\
    111 & 2008 nx & 2.29 & 2.29 & 2.38 & 2.32 & 2.80 & 2022-Apr-04 & 25.1 & 25-56 & 56 & 1 \\
    112 & (99942) apophis & 2.03 & 2.93 & 2.79 & 2.87 & 2.80 & 2020-Jun-30 & 19.2 & 384-859 & 4455 & 10 \\
    113 & 2017 bl30 & 2.37 & 3.24 & 2.79 & 2.79 & 2.85 & 2022-Apr-04 & 23.3 & 58-130 & 319 & 1 \\
    114 & 2014 lk21 & 2.11 & 2.90 & 2.79 & 2.87 & 2.91 & 2023-Oct-05 & 25.9 & 17-39 & 44 & 1 \\
    115 & 2016 hl3 & 1.88 & 2.11 & 2.50 & 3.34 & 2.94 & 2021-Oct-01 & 24.7 & 30-68 & 41 & 1 \\
    116 & (65679) 1989 uq & 2.42 & 2.79 & 2.66 & 2.67 & 2.94 & 2023-Apr-04 & 19.4 & 350-783 & 399 & 10 \\
    117 & 2008 lg2 & 2.32 & 2.67 & 2.83 & 2.84 & 2.95 & 2020-Sep-05 & 25.2 & 24-54 & 108 & 2 \\
    118 & 2011 yu74 & 2.49 & 2.81 & 2.73 & 3.15 & 2.96 & 2023-Jun-18 & 23.0 & 66-149 & 109 & 3 \\
    119 & 2010 jk1 & 1.29 & 1.80 & 2.76 & 2.46 & 2.96 & 2020-Jun-17 & 24.4 & 35-78 & 79 & 4 \\
    120 & 2014 kf39 & 1.61 & 2.32 & 2.87 & 2.71 & 3.00 & 2020-Jun-02 & 25.3 & 23-51 & 51 & 2 \\
    121 & 2015 mc & 2.36 & 3.37 & 2.71 & 2.85 & 3.01 & 2021-Apr-03 & 24.1 & 40-89 & 102 & 1 \\
    122 & (471984) 2013 ue3 & 2.31 & 2.84 & 2.57 & 2.75 & 3.03 & 2020-Apr-03 & 22.8 & 73-163 & 114 & 4 \\
    123 & 2010 fv9 & 2.47 & 2.62 & 2.58 & 2.72 & 3.03 & 2023-Sep-23 & 25.4 & 22-49 & 71 & 2 \\
    124 & 2016 wt7 & 2.26 & 2.92 & 2.78 & 2.89 & 3.04 & 2020-Aug-01 & 25.9 & 17-39 & 61 & 1 \\
    125 & 2006 gb1 & 2.49 & 3.18 & 2.94 & 2.91 & 3.08 & 2023-Apr-27 & 23.6 & 50-113 & 42 & 1 \\
    126 & 2017 ed3 & 2.47 & 2.90 & 2.95 & 3.09 & 3.08 & 2022-Jun-21 & 24.2 & 38-85 & 50 & 1 \\
    127 & 2005 ka & 2.18 & 2.33 & 2.95 & 2.77 & 3.11 & 2020-May-21 & 24.7 & 30-68 & 62 & 2 \\
    128 & 2000 ag6 & 1.89 & 2.58 & 3.31 & 3.22 & 3.12 & 2023-Apr-18 & 25.3 & 23-51 & 68 & 1 \\
    129 & 2015 vo105 & 2.02 & 2.54 & 3.05 & 3.05 & 3.12 & 2023-Oct-05 & 24.1 & 40-89 & 153 & 2 \\
    130 & 2006 qq56 & 1.02 & 2.26 & 2.84 & 3.35 & 3.13 & 2021-Apr-03 & 25.9 & 17-39 & 86 & 1 \\
    131 & 2011 es4 & 2.46 & 3.23 & 2.95 & 3.16 & 3.18 & 2020-Apr-04 & 25.7 & 19-43 & 44 & 1 \\
    132 & 2017 hu2 & 2.48 & 2.60 & 3.22 & 3.33 & 3.26 & 2023-Jul-09 & 25.9 & 17-39 & 97 & 1 \\
    133 & 2006 wb & 2.35 & 2.61 & 3.22 & 3.37 & 3.37 & 2023-May-05 & 22.8 & 73-163 & 73 & 2 \\
    134 & 2009 sh2 & 2.44 & 2.66 & 3.14 & 3.51 & 3.38 & 2021-Apr-03 & 24.9 & 27-62 & 107 & 1 \\
    135 & 2006 dn & 2.47 & 2.58 & 2.83 & 3.04 & 3.38 & 2023-Apr-04 & 24.5 & 33-74 & 150 & 1 \\
    136 & 2016 co29 & 2.36 & 3.02 & 3.29 & 3.25 & 3.39 & 2023-Apr-14 & 25.0 & 26-59 & 72 & 1 \\
    137 & 2017 fr & 2.50 & 3.14 & 3.01 & 3.69 & 3.40 & 2023-Oct-05 & 25.2 & 24-54 & 80 & 1 \\
    138 & 2013 py38 & 2.34 & 2.88 & 3.18 & 3.37 & 3.52 & 2023-Oct-05 & 24.7 & 30-68 & 46 & 2 \\
    139 & 2010 te & 2.49 & 3.14 & 3.16 & 3.08 & 3.58 & 2023-Apr-19 & 26.0 & 16-37 & 44 & 1 \\
    140 & 2016 wg7 & 2.45 & 3.19 & 3.41 & 3.23 & 3.59 & 2023-Apr-04 & 26.0 & 16-37 & 124 & 1 \\
    141 & 2002 vx91 & 1.95 & 2.87 & 3.77 & 3.71 & 3.68 & 2021-Jun-05 & 24.3 & 36-82 & 43 & 3 \\
    142 & 2015 rt1 & 2.34 & 2.83 & 3.02 & 3.25 & 3.90 & 2020-Apr-03 & 25.4 & 22-49 & 74 & 1 \\
    143 & 2013 yg & 2.44 & 2.93 & 3.87 & 3.79 & 4.03 & 2020-Jun-24 & 25.4 & 22-49 & 46 & 1 \\
    \bottomrule
\end{longtable}
\endgroup
\vspace {-6mm}

\section{Conclusion}
\noindent When performing target selection during the preliminary design phases of an interplanetary mission the details on the dynamics considered plays an important role. We quantified such a role by studying the case of a small interplanetary CubeSat (M-ARGO) able to reach the vicinity of the Earth orbit with its solar electric propulsion system. We perform and compare the target selection and ranking using five different and increasingly complex models and find that the use of a simple low-thrust model improves considerably upon impulsive models reducing the probability of mis-ranking asteroid pairs from 19\% to 8\% and the number of wrongly classified asteroids (in the reachable / non reachable classes) from 30 to 15. We also find that the contribution of the Earth's own gravity is non-significant for the purpose of target selection and ranking despite the launch from the Sun-Earth $L_2$ Lagrangian point considered in the M-ARGO study case.

\section*{Acknowledgments}
We would like to acknowledge and thank Dr. Roger Walker for leading the team during the Concurrent Design Facility (CDF) \mbox{M-ARGO} study, our colleagues in ESA/ESOC who motivated us to perform this work and all the CDF study members without whose contribution the M-ARGO study case here used would not have existed.

\bibliographystyle{astrobib}
\bibliography{refs}

\end{document}